\documentclass[twocolumn,twoside,letterpaper,11pt]{article}
\usepackage{graphicx,xrrc,natbib}
\usepackage{times}
\begin{document}
\def\aj{AJ}%
\def\araa{ARA\&A}%
\def\apj{ApJ}%
\def\apjl{ApJ}%
\def\apjs{ApJS}%
\def\ao{Appl.~Opt.}%
\def\apss{Ap\&SS}%
\def\aap{A\&A}%
\def\aapr{A\&A~Rev.}%
\def\aaps{A\&AS}%
\def\azh{AZh}%
\def\baas{BAAS}%
\def\jrasc{JRASC}%
\def\memras{MmRAS}%
\def\mnras{MNRAS}%
\def\pra{Phys.~Rev.~A}%
\def\prb{Phys.~Rev.~B}%
\def\prc{Phys.~Rev.~C}%
\def\prd{Phys.~Rev.~D}%
\def\pre{Phys.~Rev.~E}%
\def\prl{Phys.~Rev.~Lett.}%
\def\pasp{PASP}%
\def\pasj{PASJ}%
\def\qjras{QJRAS}%
\def\skytel{S\&T}%
\def\solphys{Sol.~Phys.}%
\def\sovast{Soviet~Ast.}%
\def\ssr{Space~Sci.~Rev.}%
\def\zap{ZAp}%
\def\nat{Nature}%
\def\iaucirc{IAU~Circ.}%
\def\aplett{Astrophys.~Lett.}%
\def\apspr{Astrophys.~Space~Phys.~Res.}%
\def\bain{Bull.~Astron.~Inst.~Netherlands}%
\def\fcp{Fund.~Cosmic~Phys.}%
\def\gca{Geochim.~Cosmochim.~Acta}%
\def\grl{Geophys.~Res.~Lett.}%
\def\jcp{J.~Chem.~Phys.}%
\def\jgr{J.~Geophys.~Res.}%
\def\jqsrt{J.~Quant.~Spec.~Radiat.~Transf.}%
\def\memsai{Mem.~Soc.~Astron.~Italiana}%
\def\nphysa{Nucl.~Phys.~A}%
\def\physrep{Phys.~Rep.}%
\def\physscr{Phys.~Scr}%
\def\planss{Planet.~Space~Sci.}%
\def\procspie{Proc.~SPIE}%
\let\astap=\aap
\let\apjlett=\apjl
\let\apjsupp=\apjs
\let\applopt=\ao


\title{Outflows and jets from massive star-forming clusters}


%
\author{H. Beuther                        } 
\institute{Harvard-Smithsonian Center for Astrophysics   } 
\address{60 Garden Street, Cambridge, MA 02138, U.S.A.   } 
\email{     hbeuther@cfa.harvard.edu                     } 
%
%
%

\maketitle

\abstract{Studying outflows from young massive star-forming clusters
allows one to deduce physical processes that lead to the formation of the
most massive stars. I will review the current state of
high-spatial-resolution interferometric (sub-)mm studies of massive
molecular outflows and their implications for high-mass star
formation. A possible evolutionary scenario for massive outflows
with highly collimated structures at the very beginning and more
wind-like outflows at later stages will be outlined.}

\section{Introduction}

Most of the subjects and contributions to this conference focus on the
radio and/or Xray regime of the electromagnetic spectrum. While these
two regimes are the extremes from the observational point of view,
many processes are best observed between these frequencies. The youngest
massive star-forming regions are typically cold and dominated by thermal
emission at temperatures between 10 and 100\,K which peak at mm/sub-mm
wavelengths. Therefore, the study of jets and outflows from such
regions has to concentrate on this wavelength regime, and regarding
the title of the conference I will focus on the {\it ``connection''}.

One of the best known outflows in massive star formation is the
chaotic outflow system emanating from the Orion-KL
region. Morphologically, it appears very different from typical
collimated low-mass outflows because it does not show any preferred
direction and resembles an explosion-like scenario (see, e.g.,
near-infrared images by \citealt{schultz1999}). It is important to
stress that this well known outflow is far from being typical, and
studies of other massive outflows --~as discussed below~-- indicate
that the Orion outflow might even be the exception. In the following,
I will concentrate on outflows and jets of massive star-forming
regions at early evolutionary stages prior or at the 
beginning of forming an ultracompact H{\sc ii} (UCH{\sc ii}) region.

Massive star formation proceeds in a clustered mode, and due to large
typical distances of a few kpc it is difficult to resolve the cluster
centers spatially in the (sub-)mm regime (see, e.g., various reviews
in Protostars and Planets IV). Even with the most advanced
interferometric imaging techniques we rarely exceed a linear spatial
resolution of a few 1000\,AU. An alternative approach to study the
formation of massive stars is to investigate the properties of the
molecular outflows and jets. These outflows take place on parsec
scales, they are far easier to resolve spatially and indirectly give
information about the physical processes taking place at the cluster
centers.

Mainly two scenarios compete to explain the formation of massive
stars: on the one hand, the low-mass paradigm is extended to high-mass
stars, and the most massive objects can form via similar
accretion-based processes as their low-mass counterparts, just with
significantly enhanced accretion rates (e.g.,
\citealt{wolfire1987,jijina1996,mckee2003,yorke2002}). On the other
hand, it was proposed that massive stars could also form at the center
of very dense clusters by mergers of intermediate-mass protostars
(e.g., \citealt{bonnell1998,stahler2000,bally2002}). Outflow studies
can help to discriminate between the two scenarios because the
accretion-based formation requires massive outflows as collimated as
their low-mass counterparts, whereas the coalescence scenario predicts
outflows to be far less collimated.

\section{Single-dish studies}

Most single-dish studies of massive molecular outflows agree that they
are ubiquitous phenomena in high-mass star formation and that they are
far more massive and energetic than their low-mass counterparts
\citep{shepherd1996a,richer2000,ridge2001,zhang2001,beuther2002b}.
However, early single-dish studies of massive outflows indicated that
they might be less collimated than low-mass outflows
\citep{shepherd1996b,ridge2001}. This led to the conclusion that
different physical processes might entrain the molecular gas, e.g.,
the deflection of infalling gas from the massive protostar
\citep{churchwell1999}. Furthermore, these results were interpreted as
support for the coalescence scenario. 

Contrary to this, more recent single-dish observations by
\citet{beuther2002b} with an angular resolution up to 5 times better
than previous studies show that the previously observed lower
collimation of massive outflows might only be an observational
artifact due to the larger distances of the outflows. Their sample of
21 bipolar outflows is consistent with massive outflows being as
collimated as low-mass flows, and thus support the accretion
scenario.

Obviously, single-dish studies allow ambiguous interpretations and only
high-spatial resolution interferometric studies of massive outflows
can shed light on the underlying outflow and star formation processes.

\section{High-spatial-resolution studies}

Over the last few years interferometric studies of massive outflows
developed a more detailed picture of their properties. Figure
\ref{05358} shows as an example the various molecular outflows toward
the young high-mass protostellar object (HMPO) IRAS\,05358+3543 that
is in an evolutionary stage prior to forming an UCH{\sc ii}
region \citep{beuther2002d}.

\begin{figure*}[htb]
\caption{\footnotesize Presented are the PdBI observations as contour
overlays on the grey-scale H$_2$ data \citep{beuther2002d}. The blue
and red lines in Figures {\bf (a) \& (b)} show the blue and red wing
emission of the CO(1--0) and SiO(2--1) emission, respectively. The
green lines in Figure {\bf (c)} present the integrated
H$^{13}$CO$^+$(1--0) emission. The numbers in brackets label the three
H$^{13}$CO$^+$ sources and the beams are shown at the bottom right. In
all images the arrows and ellipses sketch the three outflows (the
arrows to the right and the ellipses represent two slightly different
interpretations of the western outflow). The three squares represent
the three mm sources, the diamonds locate the H$^{13}$CO$^+$ peaks,
and the triangle marks the IRAS 12\,$\mu$m position. The inlay at the
top-left shows a close-up of the central 3 mm sources (grey-scale) and
the high-velocity outflow.}
\label{05358}
\end{figure*}

The main feature in Figure \ref{05358}(a) is the highly collimated
molecular CO(1--0) outflow emanating from a 100\,M$_{\odot}$ dust
condensation and terminating in bow-shocks observed in H$_2$
emission. The collimation degree (length divided by width) of this
outflow is 10 --~as high as the highest values reported for low-mass
flows \citep{richer2000}~--, and the mass of entrained gas is $\sim
10$\,M$_{\odot}$. From the outflow rate we can estimate an accretion
rate for the central object of the order a few times
$10^{-4}$\,M$_{\odot}$/yr. About $30''$ to the west, we observe a second
outflow which is easier depicted in the SiO(2--1) emission in Figure
\ref{05358}(b). For this flow, we do not detect a driving source in
the mm continuum (mass sensitivity $\sim 50$\,M$_{\odot}$), but the
simultaneously observed H$^{13}$CO$^+$(1--0) data show two peaks near
the outflow center, one of which likely harbors the driving source.
In addition to these two outflows, we detect a third high-velocity gas
outflow emanating again from the main 100\,M$_{\odot}$ dust
condensation at a position angle of 45$^{\circ}$ with respect to the
collimated large-scale outflow (see the inlay in Figure
\ref{05358}). A detailed analysis of the whole region can be found in
\citet{beuther2002d}.

The bolometric luminosity of this source is $10^{3.8}$\,L$_{\odot}$
corresponding to a B1 star, thus we have not reached the regime of
genuine O stars yet. Nevertheless, the data show that very young stars
greater 10\,M$_{\odot}$ can exhibit massive and collimated outflows,
and the estimated accretion rate is consistent with the
accretion-based formation of massive stars (e.g.,
\citealt{norberg2000,mckee2003}). Contrary to this, it is hard to
imagine how such a collimated structure could survive during the
highly energetic and eruptive processes of potential protostellar
mergers. The overall picture of this region appears to be
complicated due to the clustered mode of formation, several
protostellar condensations and outflows, but with high enough spatial
resolution we can disentangle the region into features well known from
low-mass star formation, new physical processes like protostellar
mergers are not necessary. Similar results were obtained via mm
interferometer studies by, e.g., \citet{beuther2003a,gibb2003,su2004};
Wyrowski et al. (in prep.). Tackling the problem from the
near-infrared, some groups started studying the shocked H$_2$ emission
at 2.1\,$\mu$m, and, e.g., Davis et al. (in prep.) also conclude from
their H$_2$ observations that the molecular jets they observe are
scaled-up versions of their low-mass counterparts.

However, not all observed massive outflows fit as well into this
scenario. For example, \citet{shepherd1998} observed the massive
outflow toward the UCH{\sc ii} region G192.16, and they find an
opening angle of the outflow of 60$^{\circ}$ (Fig. \ref{g192}). This
outflow is consistent with both -- the presence of a poorly collimated
wind and a jet \citep{shepherd1998}. Additional studies of outflows
toward UCH{\sc ii} regions in W75 \citep{shepherd2003} also indicate
that these massive outflows are not just scaled-up version of low-mass
flows.

\begin{figure*}[htb]
\caption{\footnotesize The blue and red CO(1--0) emission toward
G192.16 observed with OVRO \citep{shepherd1998}. The inlay shows a
close-up of the central mm continuum driving source. The synthesized
beams are shown at the bottom left/right of each panel.}
\label{g192}
\end{figure*}

{\bf Position-velocity (p-v) diagrams:} In addition to morphological
interpretations of massive outflows it is important to study their
kinematics as well. \citet{beuther2004d} compared the p-v diagrams of
four massive outflows, ranging in luminosity from intermediate-mass
protostars to high-mass regions with $10^{4.9}$\,L$_{\odot}$, with
previous studies in the low-mass regime by
\citet{lee2000,lee2002}. The various p-v diagrams are shown in Figure
\ref{p-v}, and we can discern various features: in some sources, we
detect high-velocity gas at the center (IRAS\,19217, IRAS\,20293,
IRAS\,23033), whereas we detect high-velocity gas at some distance
from the outflow center in all regions but with various signatures:
IRAS\,23033 exhibits the well-known Hubble-law, i.e. a velocity
increase with distance from the outflow center; in IRAS\,20293 the end
of the outflow shows emission at all velocities; and IRAS\,19217 shows
some high-velocity features but without any real symmetry with regard
to the core center; the p-v digram toward IRAS\,20126 is very
symmetric showing first increasing and then decreasing velocities with
distance from the core center.

\begin{figure*}[htb] 
\caption{\footnotesize Position-velocity diagrams of the four
intermediate- to high-mass outflows presented in
\citet{beuther2004d}. The horizontal lines mark the centers of the
outflows which always correspond to the main mm continuum sources. At
the top left of each panel we show the resolution
element. \label{p-v}}
\end{figure*}

The variety of different features in the p-v plane is broad,
but comparing them to low-mass studies, the same features appear in
p-v diagrams for sources of all
masses. \citet{lee2000,lee2001,lee2002} model the different features
by wind-driven outflows on the one hand and jet-driven outflows on the
other hand. No single model can reproduce all observations
consistently. Therefore, they conclude that outflows can be driven by
both processes, in most sources one or the other process is dominant
but sometimes wind- and jet-driven contributions can be observed
within the same outflow.

Apparently, this statement also holds for high-mass stars, we observe
jet-like morphologies and corresponding features in the p-v domain as
well as wind-like less collimated outflows and their corresponding p-v
signatures.

{\bf Evolutionary sequence?} The statistical number of observed massive
outflows with sufficiently high angular resolution is still poor, but
considering the evolutionary state of different sources some tentative
interpretations are possible. The most collimated jet-like massive
outflows have been observed toward the earliest stages of massive star
formation, the so called HMPOs which have not yet formed any
UCH{\sc ii} region (e.g., IRAS\,05358+3543), whereas the
less collimated and more wind-like outflows were detected toward
slightly more evolved UCH{\sc ii} regions (e.g.,
W75). This evolutionary difference triggers the speculation that
the various observed outflow morphologies and p-v diagrams could be
due to the evolution of the underlying massive protostar.

In this scenario, at the earliest stages massive protostars should be
surrounded by massive accretion disks and drive collimated jet-like
outflows as observed toward their low-mass counterparts. Shortly after
the central protostar ignites, we observe an UCH{\sc ii}
region and the stellar wind contributes a less collimated component to
the outflow. During the following evolution, the disk and the jet-like
outflow are unlikely to survive and only the wide-angle wind-like
outflow remains observable.

As the evolutionary timescales get shorter the more massive the stars
are, it is likely that the timescale for jet-like outflows in evolving
early O stars should be extremely short, whereas collimated outflows
should be observable significantly longer toward early B stars. This
could also explain that we do observe collimated outflows toward B
stars like IRAS\,05358+3543 whereas we were not yet successful in
finding any toward genuine early O stars. In this scenario, it will be
highly difficult to observe collimated outflows toward early O stars
just because they are rare and the corresponding evolutionary stage so
short-lived.

Furthermore, during the formation of an O star, the protostellar
region exhibits for some time {\it only} the luminosity of a B star,
and that could be just the time where we can observe the collimated
outflows. Accreting further and reaching the luminosity of an O star
the radiation and wind of the evolving central driving source can
already be so strong that it dominates any previously observable
collimated structures. In that framework, it could be intrinsically
impossible to ever observe a collimated outflow from an O star because
collimated structures would only exist in the phase when the source
has not reached its O star luminosity yet.

\section{The Xray and radio regime}

As mentioned in the introduction, this contribution focuses' mainly on
the (sub-)mm regime, thus the {\it ``connection''} of the Xray and
radio wavelengths. Nevertheless, I will shortly discuss what we can
learn from Xray and radio observations of massive outflows and jets.

{\bf Xray:} Regarding Xray observations, the discussion remains
short because so far there have been no detections of any outflow
features in the Xray regime toward very young massive star-forming
regions. There are studies toward various regions of massive star
formation (Orion, \citealt{garmire2000}; W3, \citealt{hofner2002};
IRAS\,19410+2336, \citealt{beuther2002e}), and many point sources were
detected in each of the fields, but no emission clearly associated
with molecular outflows or jets. The situation is slightly different
for more evolved massive star-forming regions, and
\citet{townsley2003} detect diffuse soft Xray emission toward a number
of regions. This emission likely arises from fast O star winds
thermalized either by wind-wind collisions or by a termination shock
against the surrounding media. More details can be found in this
volume's contribution by Leisa Townsley. At the low-mass end of
molecular outflows there are few detections of Xray emission
toward bow-shocks (e.g., in L1551, \citealt{favata2002}) but these are
also rare.

{\bf Radio:} Contrary to the Xray-regime, there are numerous
observations of radio jets toward massive star-forming regions. Figure
\ref{radio} shows two examples, the left one a typical thermal radio
jet, and to the right the unique case of a synchrotron jet in star
formation.

\begin{figure*}[htb] 
\caption{\footnotesize The left panel shows the thermal radio jet
observed toward Cepheus A HW2, the crosses mark H$_2$O maser positions
\citep{torrelles1996}. In the right panel we show the radio
synchrotron jet observed toward W3H$_2$O, the arrows outline proper
motion directions of H$_2$ masers in this field \citep{wilner1999}
\label{radio}}
\end{figure*}

The most commonly observed radio jets have a positive spectral index
due to a partially optically thick thermal free-free emission
\citep{anglada1998}. The spatial extent of these radio jets is always
very small -- a few arc-seconds -- compared to the large
scale molecular outflows discussed in the previous section. These
radio jets are likely to be at the base of the large-scale outflows
observed at (sub-)mm wavelengths. 

While thermal radio jets are typical, the radio jet in W3(H$_2$O)
exhibits a negative spectral index \citep{reid1995} consistent with
synchrotron emission. While synchrotron jets are observed regularly
toward external galaxies, this is the only known synchrotron jet
emanating from a star-forming region. The 1.6\,GHz flux of 2.5\,mJy
can be extrapolated to the Xray-regime ($S\propto \nu^{-0.6}$) and the
expected flux at 3\,keV would be 0.02\,$\mu$Jy$\sim 2\times
10^{-23}$\,erg\,cm$^{-2}$, too weak to be detected by CHANDRA or XMM.
Recently, \citet{shchekinov2004} modeled the synchrotron emission of
W3H$_2$O via the interaction of a stellar wind with the surface of a
circumstellar accretion disk.

An interesting but likely circumstantial difference between the
thermal radio jet in Cepheus A and the synchrotron jet in W3(H$_2$O) is
the associated H$_2$O maser emission. While the H$_2$O masers in
Cepheus A are perpendicular to the jet orientation and thus probably
associated with an accretion disk \citep{torrelles1996}, the proper
motions of the H$_2$O maser in W3(H$_2$O) indicate that the maser
emission traces the jet as well. For more details on H$_2$O masers and
their origin I refer to the reviews by \citet{garay1999,kylafis1999}.

\section{Discussion}

While we can study thermal jets in young massive
star-forming regions at radio wavelengths, the Xray-regime appears to
be far less promising. However, the most interesting wavelength band
for studies of molecular jets and outflows is the (sub-)mm band.

Recent years have brought tremendous progress, and in addition to the
fact that massive outflows are ubiquitous phenomena, we now know far
more details about the masses, kinematics and energetics. There has
been a lot of discussion as to whether massive outflows are less
collimated than their low-mass counterparts or
not. High-spatial-resolution observations of various massive outflows
have shown that very collimated jet-like structures do exist at the
earliest evolutionary stages of high-mass star formation. Furthermore,
comparisons of position-velocity diagrams with data from low-mass
sources show that we observe similar kinematic structures for outflows
of all masses, and that jet-driven as well as wind-driven models are
necessary to explain all features. The statistical database of
high-angular-resolution observations is still pretty poor, but the
data are suggestive of a scenario where the collimation of massive
outflows changes with time: at the earliest stages prior to forming an
UCH{\sc ii} region collimated jet-like outflows appear to be
present. As soon as the central protostar ignites and starts to form
an UCH{\sc ii} region, wind-driven processes should come into play and
contribute to the observed less collimated outflow
morphology. Evolving further, the radiation of the central protostar
starts destroying the accretion disk and collimated structures cannot
survive anymore. In this scenario, it is no surprise that we did not
detect a collimated outflow from a genuine O protostar yet because
their evolutionary timescales are extremely short and it will be
difficult to find and identify any such source.

\section*{Acknowledgments}
I like to thank Andrew Walsh for critically reviewing the initial
manuscript. I acknowledge financial support by the
Emmy-Noether-Program of the Deutsche Forschungsgemeinschaft (DFG,
grant BE2578/1).

\end{document}